\documentclass[twocolumn,english,aps,prl,superscriptaddress]{revtex4}
\usepackage[T1]{fontenc}
\usepackage[latin9]{inputenc}
\usepackage{xcolor}
\usepackage{pdfcolmk}
\usepackage{amsmath}
\usepackage{amssymb}
\usepackage{graphicx}
\usepackage{esint}
\PassOptionsToPackage{normalem}{ulem}
\usepackage{ulem}

\makeatletter

\providecolor{lyxadded}{rgb}{0,0,1}
\providecolor{lyxdeleted}{rgb}{1,0,0}
\@ifundefined{textcolor}{}
{%
 \definecolor{BLACK}{gray}{0}
 \definecolor{WHITE}{gray}{1}
 \definecolor{RED}{rgb}{1,0,0}
 \definecolor{GREEN}{rgb}{0,1,0}
 \definecolor{BLUE}{rgb}{0,0,1}
 \definecolor{CYAN}{cmyk}{1,0,0,0}
 \definecolor{MAGENTA}{cmyk}{0,1,0,0}
 \definecolor{YELLOW}{cmyk}{0,0,1,0}
}

\makeatother

\usepackage{babel}
\begin{document}
\global\long\def\oc#1{\hat{c}_{#1}}
 \global\long\def\ocd#1{\hat{c}_{#1}^{\dagger}}
 \global\long\def\tr{\text{Tr}\,}
 \global\long\def\im{\text{Im}\,}
 \global\long\def\re{\text{Re}\,}
 \global\long\def\bra#1{\left\langle #1\right|}
 \global\long\def\ket#1{\left|#1\right\rangle }
 \global\long\def\braket#1#2{\left.\left\langle #1\right|#2\right\rangle }
 \global\long\def\obracket#1#2#3{\left\langle #1\right|#2\left|#3\right\rangle }
 \global\long\def\proj#1#2{\left.\left.\left|#1\right\rangle \right\langle #2\right|}

\title{Absence of Diffusion in an Interacting System of Spinless Fermions
on a One-dimensional disordered lattice}

\author{Yevgeny Bar Lev}

\affiliation{Department of Chemistry, Columbia University, New York, New York
10027, USA}

\email{yb2296@columbia.edu}

\selectlanguage{english}%

\author{Guy Cohen}

\affiliation{Department of Chemistry, Columbia University, New York, New York
10027, USA}

\affiliation{Department of Physics, Columbia University, New York, New York 10027,
USA}

\author{David R. Reichman}

\affiliation{Department of Chemistry, Columbia University, New York, New York
10027, USA}
\begin{abstract}
We study the infinite temperature dynamics of a prototypical one-dimensional
system expected to exhibit many-body localization. Using numerically
exact methods, we establish the dynamical phase diagram of this system
based on the statistics of its eigenvalues and its dynamical behavior.
We show that the nonergodic phase is reentrant as a function of the
interaction strength, illustrating that localization can be reinforced
by sufficiently strong interactions even at infinite temperature.
Surprisingly, within the accessible time range, the ergodic phase
shows subdiffusive behavior, suggesting that the diffusion coefficient
vanishes throughout much of the phase diagram in the thermodynamic
limit. Our findings strongly suggest that Wigner\textendash Dyson
statistics of eigenvalue spacings may appear in a class of ergodic
but subdiffusive systems.
\end{abstract}
\maketitle
The interplay between particle interactions and disorder may lead
to complex emergent phenomena, especially in one-dimensional systems
where the influence of both effects is maximized. Non-interacting
particles in a one-dimensional disordered system exhibit Anderson
localization \cite{Anderson1958b} which results in insulating, nonergodic
behavior. Coupling the localized system to phonons will restore ergodicity
and transport with a peculiar dependence on the temperature, a phenomenon
know as variable-range hopping \cite{Mott1969}. In the absence of
phonons or coupling to any other degrees of freedom it was generally
believed that the inter-particle interactions conspire to induce transport
and restore ergodicity \cite{Fleishman1978}, although the opposite
was also suggested in a later study \cite{Fleishman1980a}. Nevertheless,
using self-consistent perturbation theory in the interaction term,
it has recently been suggested that the localized phase survives finite
interactions \cite{Gornyi2005,Basko2006a}. Moreover, \emph{the many-body}
spectrum is predicted to have a mobility edge separating the localized
and metallic states, similar to the one-particle mobility edge in
three-dimensional systems \cite{Mott1967}. For lattice models where
the energy density is bounded, it has been proposed that a range of
parameters might exists for which \emph{all} the many-body eigenstates
are localized, such that the many-body localization (MBL) transition
persists at infinite temperatures \cite{Oganesyan2007a}. This argument
has recently been made more precise \cite{BarLev2014,Ros2014}. The
existence of a nonergodic phase for strong disorder and weak interactions
has been rigorously proven for zero particle density \cite{Aizenman2009b}
and for an infinite chain of spins \cite{Imbrie2014}. However, currently
there are no rigorous results for the ergodic phase or the MBL transition
itself. Although realizing a truly isolated physical system is impossible,
recent experiments in cold atom systems come very close to this idealized
limit \cite{Bloch2008,Kondov2013,Senko2014}.

The MBL transition is a dynamical transition between a nonergodic
and an ergodic phase, and it has no manifestation in static thermodynamic
quantities. Its unconventional nature has attracted many researchers.
In particular, the dynamical features of the transition which have
been studied are the dc conductivity \cite{Karahalios2009a,Barisic2010a,Berkelbach2010a}
and dynamical correlations in the $t\to\infty$ limit \cite{Pal2010a}.
In all of these studies exact diagonalization (ED) has been used,
effectively restricting the accessible system sizes to about $16$
sites. This fact poses serious limitations on the interpretation of
the results. In particular, the evaluation of the dc conductivity
depends on a careful extrapolation to the thermodynamic limit \cite{Thouless1981},
while for systems of finite length $L$, dynamics in the $t\to\infty$
limit are dominated by finite size effects and may have little in
common with the behavior of $L\to\infty$ system. Another measure
which has been used to study the MBL transition is the distribution
of spacings of the many-body eigenvalues. At the transition the distribution
is expected to cross over from a Poisson to a Wigner--Dyson distribution
\cite{Oganesyan2007a,Pal2010a}. For one-particle systems, it has
been conjectured in Ref.~\cite{Bohigas1986} that quantum systems
with fully chaotic classical analogs will exhibit a Wigner--Dyson
distribution of eigenvalue spacing. It was later shown that the Wigner--Dyson
distribution of \emph{many-body} eigenvalue spacing is generic and
connected to \emph{quantum} non-integrability \cite{Montambaux1993,Poilblanc1993}.
Nevertheless, the connection between non-integrability and transport
properties is not rigorously understood. For clean, translationally
invariant systems, non-integrability generally results in the disappearance
of ballistic transport \cite{Narozhny1998,Heidrich-Meisner2005,Heidrich-Meisner2007},
however for disordered \emph{many-body} systems its implications have
not been fully explored. A number of numerically exact studies have
examined the dynamics directly. It has been shown that time-dependent
density matrix renormalization group (tDMRG) becomes efficient for
highly localized systems \cite{Znidaric2008}. For weak interactions
a logarithmic growth of entanglement entropy as a function of time
has been observed \cite{Znidaric2008,Bardarson2012} and later explained
\cite{Vosk2013a,Huse2013,Serbyn2013a}. Entanglement entropy is however
a non-local quantity with no direct relation to the measurable dynamical
properties of the system. Two of the authors in a previous work directly
observed nonergodicity by studying the relaxation of the on-site particle
density, in a study limited to weak interactions \cite{BarLev2014}.
In a very recent study quantum revivals of the local density were
used to differentiate between the Anderson and MBL localized phases
\cite{Vasseur2014}.

In this letter we explore the dynamical phase diagram of a system
of interacting spinless fermions in a one-dimensional disordered lattice
via the examination of the spectral properties and the transport of
correlations in the system. The Hamiltonian we consider is given by

\begin{eqnarray}
H & = & -t\sum_{i}\left(\hat{c}_{i}^{\dagger}\hat{c}_{i+1}+\hat{c}_{i+1}^{\dagger}\hat{c}_{i}\right)\label{eq:t-V_model}\\
 & + & V\sum_{i}\left(\hat{n}_{i}-\frac{1}{2}\right)\left(\hat{n}_{i+1}-\frac{1}{2}\right)+\sum_{i}h_{i}\left(\hat{n}_{i}-\frac{1}{2}\right),\nonumber 
\end{eqnarray}
where $t$ (which we set to one) is the hopping matrix element, $V$
is the interaction strength and $h_{i}$ are random on-site fields
independently distributed on the interval $h_{i}\in\left[-W,W\right]$.
Note that by using the Jordan--Wigner transformation, this model can
be exactly mapped onto the XXZ model. Extending the model \eqref{eq:t-V_model}
to a non-integrable (zero field) version (e.g. the model used in Ref.~\cite{Oganesyan2007a})
produces only \emph{quantitative} and not qualitative changes to our
conclusions. We therefore focus on \eqref{eq:t-V_model}. For lattice
models with a finite number of states per site, the energy density
is bounded, which renders the infinite temperatures limit meaningful.
To simplify the discussion we follow Ref.~\cite{Oganesyan2007a}
and consider only the infinite temperature limit throughout this Letter.

To establish the full dynamical phase diagram using eigenvalue statistics
we repeat the analysis of Ref.~\cite{Oganesyan2007a} for a large
set of parameters ($1\leq W\leq7$ and $0.5\leq V\leq10$, a total
of 120 points). For this purpose we obtain the eigenvalues of the
Hamiltonian \eqref{eq:t-V_model} for system sizes $L=10$, 12 and
$14$ and calculate the metric $r_{n}=\min\left(\delta_{n},\delta_{n-1}\right)/\max\left(\delta_{n},\delta_{n-1}\right)$,
where $\delta_{n}\equiv E_{n}-E_{n+1}$ is the difference between
adjacent eigenvalues. This metric is then averaged over all states
and disorder realizations (100 realizations were sampled) and is used
to differentiate between Wigner--Dyson $\left(r=0.529\right)$ and
Poisson statistics $\left(r=0.386\right)$ of the eigenvalue spacing
\cite{Oganesyan2007a}. It is assumed that the metric $r\left(W,V\right)$
flows to the Wigner--Dyson value in the thermodynamic limit for the
ergodic parts of the phase diagram, and similarly to the Poisson value
for nonergodic regions. The phase boundary will therefore correspond
to points which are ``stationary'' under scaling of system size.
Note that the phase boundaries have to be taken with care; due to
the severe limitation on the available system sizes we cannot perform
a reliable extrapolation of this procedure to the thermodynamic limit.

\begin{figure}
\begin{centering}
\includegraphics[clip,width=8.6cm]{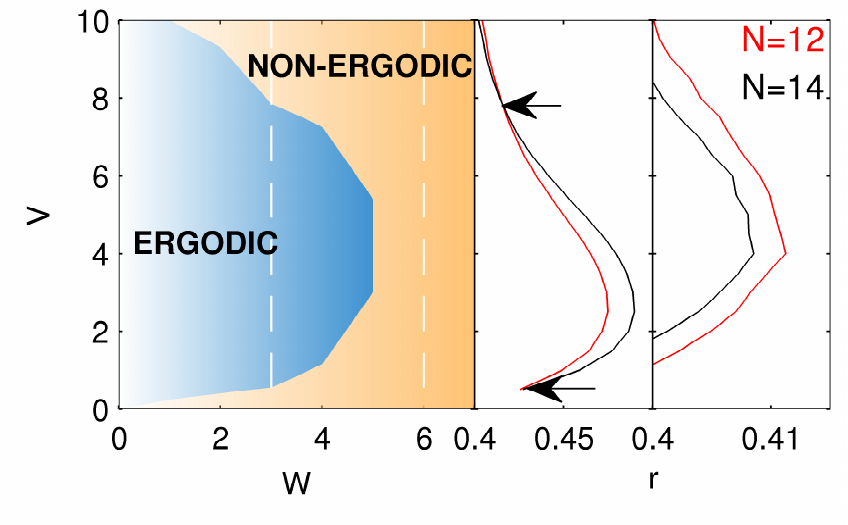} 
\par\end{centering}

\protect\caption{\label{fig:phase-diagram-stat}(Color online) Dynamical phase diagram
at infinite temperature, as obtained from the spectral fluctuations
of the studied model. The dashed white lines correspond to cuts through
the phase diagram presented on the right panel. The right panel demonstrates
the determination of the phase diagram based on spectral fluctuation
analysis of two system sizes, $N=12$ and $N=14$. The phase boundary
is determined from the crossing of the red (grey) and black lines,
as designated by the arrows. The faded region on the left indicates
a region of substantial finite size effects.}
\end{figure}

In Fig.~\ref{fig:phase-diagram-stat}, the resulting phase diagram
is presented. A surprising feature of the diagram is the re-entrant
behavior of the nonergodic glassy phase. This feature was overlooked
in previous studies, which examined only one constant interaction
cut through the diagram \cite{Oganesyan2007a,Pal2010a} or for weak
interactions \cite{Vasseur2014}. It should be noted that in Ref.~\cite{DeLuca2013a}
, a suggestion that reentrance may occur in MBL systems was put forward.
The re-entrant behavior suggests that sufficiently strong interactions
can enhance rather than destroy localization, a phenomena somewhat
reminiscent of the Mott transition occurring at low temperatures.
Note that while the \emph{clean} system is \emph{insulating} at zero
temperature for $V/t>2$ \cite{Giamarchi2004}, it exhibits \emph{diffusive}
transport at infinite temperature \cite{Steinigeweg2009,Prosen2009,Znidaric2011,Karrasch2014}.
Therefore, it is the disorder which facilitates localization.

As discussed above, Wigner--Dyson statistics of the level spacing
suggest that the system is non-integrable, but for a disordered interacting
system there are no established implications for the dynamics. Therefore,
it is interesting to examine the dynamics directly across the entire
phase diagram. For this purpose, we have used a combination of ED
and tDMRG techniques to evaluate the density-density correlation function
at infinite temperature,
\begin{equation}
C_{ij}\left(t\right)=\frac{1}{Z}\tr\delta\hat{n}_{i}\left(t\right)\delta\hat{n}_{j}\left(0\right),
\end{equation}
where $\delta\hat{n}_{i}\equiv\hat{n}_{i}-1/2$ and $Z$ is the dimension
of the Hilbert space. To eliminate boundary effects it would be preferable
to excite the system in the middle of the chain. However, to make
the best use ED, which is limited to small system sizes, we instead
use open boundary conditions and excite the system at one boundary.
This allows for the study of transport over the entire system length,
effectively increasing the accessible times. In particular, when the
excitation has traveled sufficiently far from the boundary, it is
expected that the dynamical characteristics will approach those of
the bulk and the initial position of the excitation will be irrelevant.
We have confirmed this by exciting the system from its center (data
not shown). To quantify the transport of correlations we define
\begin{equation}
\sigma^{2}\left(t\right)=\sum_{n=0}^{L-1}n^{2}\left(C_{n0}\left(t\right)-C_{n0}\left(0\right)\right).\label{eq:sigma2}
\end{equation}
This quantity measures the \emph{spreading} of correlations analogously
to the mean square displacement of a diffusing particle. A similar
quantity based on the one-time density $\left(\hat{n}_{i}\left(t\right)\right)$
has been studied extensively in clean systems out-of-equilibrium \cite{Langer2009,Karrasch2014}.
However, such quantities cannot be directly used in equilibrium where
any one-time operator is conserved. We therefore consider the spreading
of two-time correlations encoded by \eqref{eq:sigma2}\cite{Bohm1992}.
The nature of the transport is assessed by examining the finite time
dynamical exponent,
\begin{equation}
\alpha\left(t\right)\equiv\frac{\mathrm{d}\ln\sigma^{2}\left(t\right)}{\mathrm{d}\ln t},\label{eq:dynam_exp}
\end{equation}
which has values $\alpha\left(t\to\infty\right)=2$ for ballistic
transport and $\alpha\left(t\to\infty\right)=1$ for diffusive transport.
For finite systems, asymptotic time dynamics will be determined by
finite size effects such as reflections from the boundaries. Since
we are only interested in the bulk transport of correlations, we limit
the considered times to the time $t_{*}$ during which the existence
of the boundary opposite to the initial excitation has no effect on
$\sigma$. Until this horizon time, the dynamics will not depend on
the simulated system size, since the infinite temperature initial
conditions are identical for all system sizes. To determine $t_{*}$
we evaluate the spreading of correlations for different system sizes
(here $L=10$, $12$ and $14$ unless otherwise stated) for every
parameter set of the Hamiltonian. $t_{*}$ is then taken to be the
longest time up to which $L=12$ and $L=14$ exhibit the same dynamics
within the chosen accuracy.

\begin{figure}
\begin{centering}
\includegraphics[width=8.6cm]{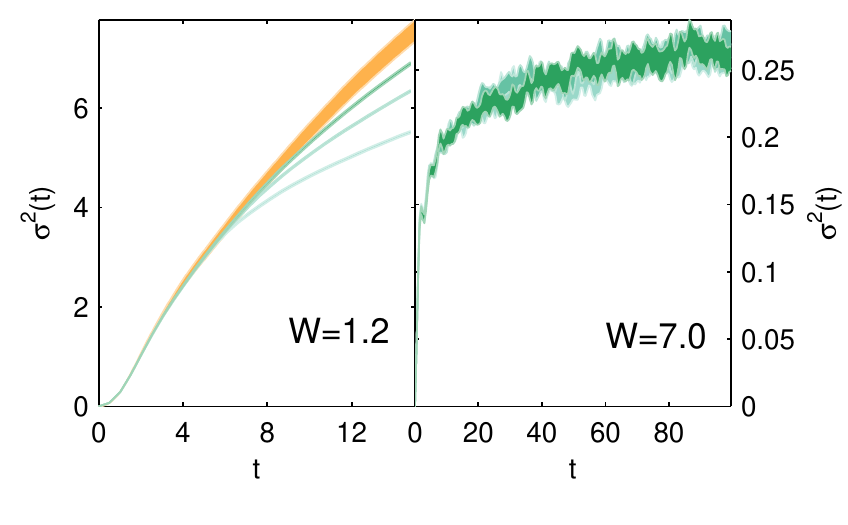} 
\par\end{centering}

\protect\caption{\label{fig:dynam-finite-size}(Color online) Finite size scaling of
the dynamical data obtained for various systems sizes at infinite
temperature. Darker colors represent larger systems. The left panel
shows $\sigma^{2}\left(t\right)$ as a function of time, for $V=2$,
and $W=1.2$ and system sizes $L=12,14$ and $16$ using ED (from
bottom) and $L=24$ using DMRG (top). The right panel shows $\sigma^{2}\left(t\right)$
for $W=7$ and system sizes $L=10,12$ and 14. Shaded areas designate
uncertainty bounds.}
\end{figure}

In Fig.~\ref{fig:dynam-finite-size} this procedure is exemplified
for two parameter sets corresponding to the ergodic and nonergodic
phases. The horizon time, $t_{*}$, is naturally much longer for the
nonergodic phase. For the chosen parameter sets it varies in the range
$5<t_{*}<100$. There are two interesting dynamical differences between
the ergodic and nonergodic phases: although similar computer time
was used in the two cases, it is clear that it is significantly harder
to converge the averaging of $\sigma^{2}\left(t\right)$ in the nonergodic
phase, as can be seen by the fluctuations of the $\sigma^{2}\left(t\right)$
in Fig.~\ref{fig:dynam-finite-size}. Another clear difference is
the appearance of oscillations in $\sigma^{2}\left(t\right)$ inside
the nonergodic phase, with a period of about $T\approx3$. This period
depends neither on the disorder strength nor the interaction, and
is related to oscillations of particles effectively localized to lattice
sites.

\begin{figure}
\centering{}\includegraphics[clip]{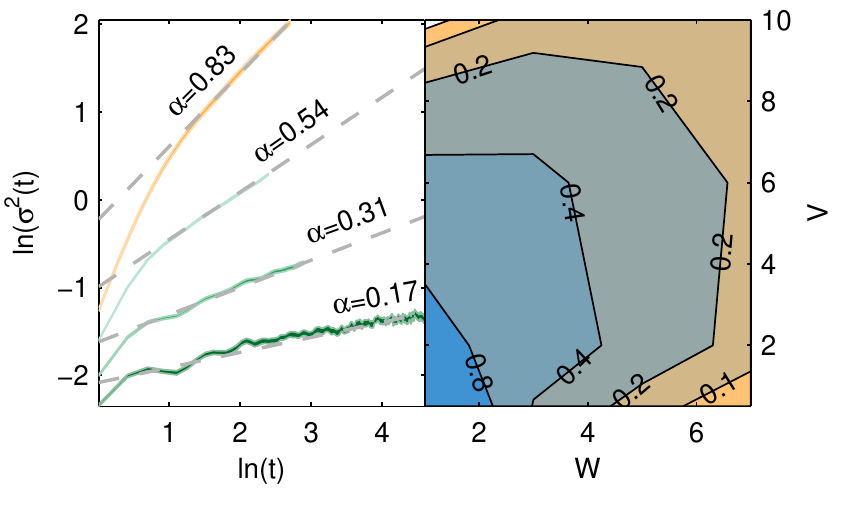} \protect\caption{\label{fig:dynam_phase}(Color online) Dynamical phase diagram obtained
from the dynamical exponents of correlation spreading. The left panel
explains the determination of the dynamical exponents from the log--log
plot of $\sigma^{2}\left(t\right)$. The system sizes which were used
$L=14$, $V=2$ and $W=3,5,7$ (from top to bottom) using ED and $L=24,$
$V=2$ and $W=1.2$ using tDMRG (top, orange). Shaded areas designate
uncertainty bounds. The dynamical exponents are extracted from the
slope of the dashed gray lines. The right panel presents the dynamical
exponents as a contour plot which interpolates between 24 different
parameter points. The text on the contour lines corresponds to the
dynamical exponents. Note that due to edge effect discussed in the
text the $W$ axis starts from $W=1$ in the right panel.}
\end{figure}

To extract the dynamical exponent $\alpha$ \eqref{eq:dynam_exp}
we first extract the horizon time $t_{*}$ for every data point in
the phase diagram, and subsequently evaluate the logarithmic derivative
of $\sigma^{2}\left(t\right)$ at $t_{*}$ using the procedure illustrated
in the left panel of Fig.~\ref{fig:dynam_phase}. Repeating the procedure
for 24 different parameter points and interpolating gives a rough
dynamical ``phase diagram'' on the right panel of Fig.~\ref{fig:dynam_phase}.
Phase boundaries cannot be reliably determined by this methodology,
since the dynamical exponents obtained are not asymptotic. For the
MBL transition scenario advocated in Ref.~\cite{Basko2006a} the
ergodic phase is diffusive, which would correspond to an asymptotic
dynamical exponent of $\alpha\left(t\to\infty\right)=1$ while the
nonergodic phase is insulating and should correspond to $\alpha\left(t\to\infty\right)=0$.
Surprisingly, the dynamical phase diagram of Fig.~\ref{fig:dynam_phase}
has a vanishingly small part with a dynamical exponent close to one.
This region corresponds to the weak localization regime, where the
\emph{non-interacting} localization length is larger than the size
of the simulated system. Interestingly, the contours of equal dynamical
exponents retrace the phase diagram of Fig.~\ref{fig:phase-diagram-stat},
exhibiting a similar re-entrant behavior. The strong localization
seen at Fig.~\ref{fig:dynam_phase} for very weak disorder and strong
interaction is an edge effect and is \emph{irrelevant} to the physics
of many-body localization. By exciting the system from its center,
using tDMRG and system size $32$ (for $W=4$ and $V=10$, data not
shown) we have verified that the re-entrant behavior is \emph{not
}influenced by this effect and is a feature that is expected to survive
extrapolation to the thermodynamic limit.

In Figs.~\ref{fig:dynam-finite-size} and \ref{fig:dynam_phase}
we used tDMRG \cite{Vidal2003,White2004,Stoudenmire2010,White2009}
to access larger system sizes wherever possible. Surprisingly, for
the purposes of this work tDMRG is superior to ED only within a narrow
parameter regime characterized by weak disorder \cite{SuppMat2014}.
For example, as illustrated in Fig.~\ref{fig:dynam_phase}, it enables
the demonstration of sub-diffusion at very weak disorder, $W=1.2$
for which the \emph{non-interacting }localization length is larger
than the system sizes accessible to us within ED.

An interesting question which remains is how the \emph{finite time}
dynamical exponents $\alpha\left(t\right)$ calculated in this Letter
change in the limit of $t\to\infty$. For classical fluids close to
the glass transition the typical scenario is slow subdiffusive transport
followed by a transition to diffusion; this, however, requires an
additional time-scale for which such acceleration of transport (de-caging)
occurs \cite{Binder2005}. For some parameter choice this time scale
can be made arbitrarily small. In our simulations we do not see the
appearance of such a time scale even for very weak disorder; throughout
the phase diagram $\alpha\left(t\right)$ (averaged over the oscillations)
is a decreasing function of time. If this is indeed the case, it implies
that transport is subdiffusive throughout the entire phase diagram.
Although for small disorder strength the validity time becomes short,
making the determination of $\alpha\left(t\right)$ less reliable,
the overall tendency of $\alpha\left(t\right)$ to decrease with time
throughout the entire phase diagram invites one to \emph{speculate}
that the small diffusive region occurring for small disorder and small
interaction will vanish in the thermodynamic limit, in the absence
of coupling to additional degrees of freedom (such as phonons). We
do not have access to the $t\to\infty$ limit, but within the attainable
times scales we can observe that $\alpha\left(t\right)$ seems to
vary continuously across the ergodic--nonergodic transition. Moreover,
although fits to logarithmic relaxation appear to be more appropriate
in the nonergodic phase, we stress that we cannot clearly distinguish
between logarithmic relaxation and weak sub-diffusion (note that logarithmic
relaxation still yields $\alpha\left(t\to\infty\right)=0$) without
accessing significantly larger timescales. While we do not show this,
logarithmic behavior is consistent with at least some of the data.
Both scenarios imply a vanishing diffusion coefficient in the thermodynamic
limit.

In summary, we have investigated the dynamical phase diagram of a
one-dimensional, spinless fermionic model with short-range interactions
and disordered potential. The phase diagram was obtained both by analysis
of the distribution of the eigenvalues spacing and the correlations
transport in the system. We showed that the nonergodic phase is re-entrant
for sufficiently strong interactions which implies that in this part
of the phase diagram disorder and interactions reinforce localization.
Moreover the phase diagram is predominantly subdiffusive for accessible
times. If this behavior persists asymptotically it implies the absence
of diffusion in the thermodynamic limit and for any finite disorder
strength. Nevertheless, the dynamical phase diagram is composed of
an ergodic, but subdiffusive phase and a nonergodic glassy phase.
Our findings imply that Wigner--Dyson statistics alone do not rule
out subdiffusive behavior. Interestingly, subdiffusive behavior was
recently experimentally observed for bosons at low temperatures \cite{Lucioni2011}.
It would be of great interest to explore the possibility of the existence
of a broad class of quantum non-integrable systems which show subdiffusive
behavior.
\begin{acknowledgments}
We would like to thank M. Schiró for many enlightening and helpful
discussions. This work used the Extreme Science and Engineering Discovery
Environment (XSEDE), which is supported by National Science Foundation
Grant No. OCI-1053575. DMRG calculations were performed using the
ITensor library, http://itensor.org. This work was supported by the
Fulbright Foundation and by Grant No. NSF-CHE-1213247.
\end{acknowledgments}

\paragraph*{Note. }

During the review process a number of works supporting the existence
of the subdiffusive phase have appeared \cite{Agarwal2014,Vosk2014,Potter2015}.
Also the strongly interacting nonergodic phase predicted in this Letter
was observed experimentally \cite{Schreiber2015a}.

\bibliographystyle{apsrev}
\bibliography{local,/home/yevgeny/Dropbox/Research/Latex/papers/Bibs/library}

\end{document}